\documentclass[twocolumn,showpacs,preprintnumbers,amsmath,amssymb,aps]{revtex4}
\usepackage{graphicx}
\begin{document}

\title{Moir\'e patterns in quantum images}
\author{J. A. O. Huguenin$^2$, M. P. Almeida$^1$, P. H. Souto Ribeiro$^1$, A. Z. Khoury$^{2*}$}
\affiliation{1- Instituto de F\'\i sica - Universidade Federal do Rio de Janeiro \\
Caixa Postal 68528, Rio de Janeiro - RJ, 21941-972, Brazil}
\affiliation{2- Instituto de F\'\i sica - Universidade Federal Fluminense \\
Niter\'{o}i - RJ, 24210-340, Brazil}

\begin{abstract}
We observed moir\'e fringes in spatial quantum correlations between twin photons 
generated by parametric down-conversion. Spatially periodic structures were nonlocally 
superposed giving rise to beat frequencies typical of moir\'e patterns. This result 
brings interesting perspectives regarding metrological applications of such a quantum 
optical setup. 
\end{abstract}

\pacs{42.50.Ar; 42.50.St; 42.50.Lc}

\maketitle

Spatial quantum correlations in light beams has attracted a great 
deal of interest lately. 
A number of recent works devoted to this subject constitute a whole new area often 
called {\it Quantum Images} \cite{misha}. 
Among the optical systems capable of experimentally producing space-time 
correlations, parametric amplifiers and oscillators are the most frequently employed. 
The striking quantum nature of the correlations created by such devices combined with 
their simplicity make them specially attractive. In cavity free parametric down-conversion, 
the conditional images are observed in the photocount regime, where the spatial correlations 
can be interpreted as a consequence of transverse momentum entanglement in the quantum state 
of the twin photons. This entanglement has been used in a variety of experiments 
\cite{monken,padua} and more recently combined with polarization entanglement to produce 
polarization controlled quantum images \cite{juliana}. 

The studies of quantum images in the high photon flux regime are complementary to the 
single photon case. In such high flux regimes, correlations in spatially dependent photocurrents 
are measured rather than photocounts. We may also quote very interesting results in this 
case, including the recent experimental demonstration of sub-shot noise measurement of 
small displacements with a spatially squeezed light beam \cite{treps}. 
The potential applications of such correlations has also motivated spatial noise measurements 
in semiconductor lasers \cite{lasers}. 

While some experimental progress has been achieved in the quantum domain, a wide 
knowledge base of classical imaging is already available. 
An attractive challenge is to match
the progress in both domains to obtain new technologies and new applications of the 
quantum optical setups. In this work we present a contribution 
involving the observation of moir\'e patterns in quantum images. When structures with 
periodic or quasi-periodic layouts are superposed, new structures appear as a 
consequence of the beat between the spatial frequencies in the spectra of the 
original layouts. These are the well known {\it moir\'e fringes} \cite{moire}. 
This effect plays an important role in many different fields like solid structures 
analysis or optical metrology, for example. 

When solid structures are studied through X-ray diffraction or 
Transmission Electron Microscopy (TEM), the superposition of 
slightly different structures gives rise to moir\'e fringes. Since the spatial beat 
frequencies are much smaller than those belonging 
to the spectra of the original structures, the moir\'e effect provides a 
considerable sensitivity enhancement to small structural imperfections 
\cite{xray,tem}. 
The moir\'e effect has also a number of applications to optical metrology. 
From surface analysis of thin films \cite{diagnostics} to strain structure 
measurements in human teeth \cite{odonto}, the moir\'e effect has been employed 
as a powerful tool for high sensitivity measurement of the spatial structure 
of surfaces. An interesting technique called {\it Projection Moir\'e 
Interferometry} consists of projecting the image of a primary periodic structure 
on a secondary one to obtain the moir\'e pattern. If the spatial frequency 
of the primary structure is previously known, then the one of the secondary 
structure can be readily obtained through the measurement of the moir\'e 
beat frequency. This method provides an excellent accuracy and has already been 
used for noncontact temperature measurements on Si with a resolution of 
$\pm 0.05^oC\,$ \cite{si}. In summary, the moir\'e technique provides high 
sensitivity to small deformations in a periodic or quasi-periodic structure. 
This property has been recently applied to an image encryption scheme 
\cite{criptografia}, where the intensity function corresponding to the image 
to be encrypted is used to deform a periodic structure. The encrypted image 
is then recovered through the moir\'e pattern between the deformed structure 
and a reference pattern. 

In our experiments, we used the spatial correlations between twin photons 
produced by spontaneous parametric down conversion to observe moir\'e patterns 
in quantum images. We exploited two important features: the image transfer from 
the pump beam to the spatial profile of the quantum correlations, and the 
combined transmission through masks remotely placed in the signal and idler 
beams. Therefore, two experimental setups were employed. In the first setup, 
one transmission grating was placed in the pump beam while the other one was 
placed in the down-converted idler beam. In the second setup, the transmission 
gratings were placed in the down-converted signal and idler beams. 

For each setup, two different regimes were investigated. In the first 
regime, the spatial periods of the gratings were quite different 
(1.2 and 1.6 mm), so that both the high and the low spatial frequencies were 
visible in the superposition of the gratings. 
In the second regime, two gratings with similar spatial periods were used 
(0.8 and 0.9 mm). In this case, only the slow beat frequency could be 
clearly followed. The standard moir\'e patterns 
obtained with the two pairs of gratings are presented in Fig. 1. They were 
registered with a CCD camera. In Figs. 1.a and 1.b we show the moir\'e patterns 
created by the first and the second pair of gratings respectively. While in 
the first case a complicated structure appears, in the second one 
the low frequency modulation can be easily identified. The dashed lines in both 
figures outline the regions scanned in the quantum optical setups.
\begin{figure}[ht]
\includegraphics[clip=,width=7cm]{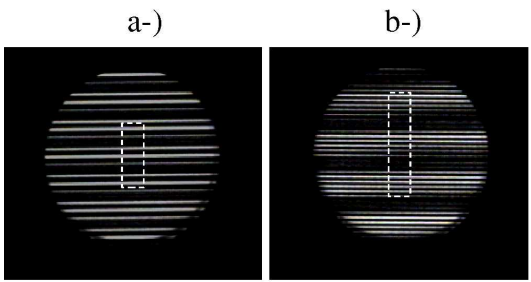}
\caption{\label{moireclass}Standard moir\'e patterns obtained with  
two pairs of gratings: a-) 1.2/1.6mm and b-) 0.8/0.9mm.}
\end{figure}
The first setup used is shown in Fig. 2. A 5 mm long LiIO$_{\rm 3}$ crystal 
was pumped by  425 nm wavelength pulses obtained from a frequency doubled 
Ti-Saf laser. 
Pairs of quantum correlated signal and idler beams were generated through 
type I parametric down-conversion at wavelengths 890 nm (signal) and 
810 nm (idler), and detected with photo-avalanche detectors (D1 and D2). 
A pinhole with 0.5 mm diameter was placed in front of each detector.
The coincidence count was performed electronically and registered 
with a computer. 
\begin{figure}[ht]
\includegraphics[clip=,width=9cm]{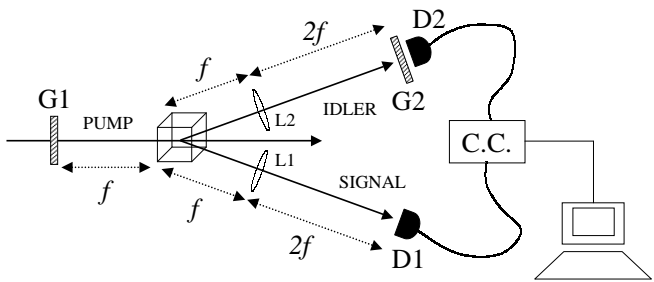}
\caption{\label{setup1}Experimental setup for pump-idler moir\'e}
\end{figure}
The first grating (G1) was placed in the pump beam before the nonlinear crystal. 
It is well known that the spatial correlations between the signal and idler beams 
propagate like the pump profile after the crystal, except for a wavelength dependent 
scaling factor \cite{monken}. The image of G1, carried by the signal-idler 
correlations, was projected on the detection plane by two identical lenses 
(L1 and L2) with focal length $f$, inserted in the down-converted beams. 
The second grating (G2) was then placed right before D2. Under such conditions, 
the spatial profile of the coincidence count is expected to be proportional to the 
product G1 $\cdot$ G2 \cite{juliana2}. 
Since down-conversion was nearly degenerate, the scaling 
factor for image transfer was close to 2. Therefore, gratings with half the 
desired spatial period were introduced in the pump beam in order to produce 
spatial correlations with the desired effective modulations. 

The coincidence region was quite narrow, which means that, for a given 
position of the signal (idler) detector, coincidences could be observed when the 
idler (signal) detector was scanned over a narrow region about 4 mm$^{\rm 2}\,$ 
in area. 
So, it was impossible to measure the quantum image by scanning the 
detectors. Instead, we fixed the detectors and scanned both gratings at the 
same time. In Fig. 3a we show the coincidence profile as a function of the 
gratings position. This profile corresponds to the moir\'e pattern shown in 
Fig. 1a. A grating with 0.8 mm period (G1) was introduced in the 
pump beam and displaced in steps of 0.1 mm, thus producing an effective 1.6 mm period 
modulation on the correlations, scanned in effective steps of 0.2 mm. 
The second grating (G2), with a 1.2 mm period, was introduced right before the 
idler detector and scanned in steps of 0.2 mm in the same sense as G1. 
For such parameters, the expected slow moir\'e modulation period is 4.8 mm, in 
very good agreement with the spacing between the large peaks in Fig.3a. 
The solid line 
in this figure is the product of two cosine squared functions with the periods 
of G1 and G2, showing a very good agreement for both the fast and the slow 
modulation. The small deviations of the experimental data from the solid 
line are probably due to distortions of the real gratings with respect to 
ideal cosine functions and to the finite size of the detectors aperture, 
which limits the spatial resolution.

The coincidence profile in Fig. 3b corresponds to the moir\'e pattern shown 
in Fig. 1b. This time, a 0.4 mm period grating (G1) was introduced in the 
pump beam, producing an effective modulation with 0.8 mm period in the 
coincidence profile. The other grating (G2) had a 0.9 mm period, therefore 
very close to that of G1. This time G1 was scanned in steps of 0.0 5mm, so 
producing an effective scan in steps of 0.1 mm, and G2 was scanned in steps 
of 0.1 mm. The high frequency modulation arising from the superposition was 
too fast, so that a meaningful fit could be provided only for the slow 
beat modulation. The solid line is a single cosine squared envelope 
with a slow beat period of 7.8 mm, which is quite close to the expected 
value of 7.2 mm.  
\begin{figure}[ht]
\includegraphics[clip=,width=8cm]{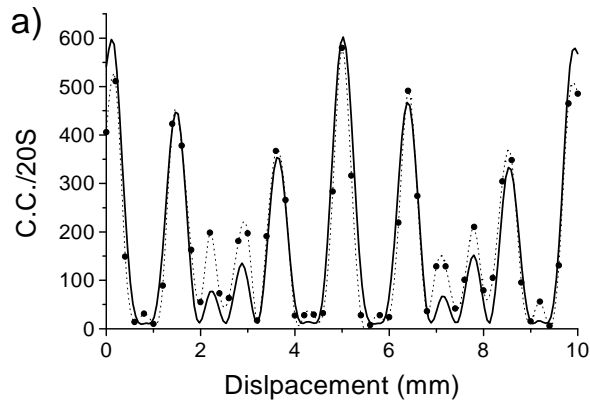}
\includegraphics[clip=,width=8cm]{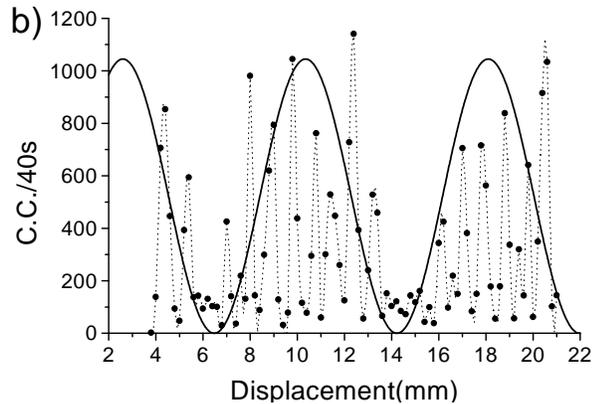}
\caption{\label{result1}Experimental results for pump-idler moir\'e.
a-) 1.2-1.6 mm moir\'e. The solid line is a 
$\cos^2(\pi\,x/1.2mm)\times\cos^2(\pi\,x/1.6mm)$ fit.
b-) 0.8-0.9 mm moir\'e. The solid line is a $\cos^2(\pi\,x/7.8mm)$ fit 
for the slow moir\'e modulation. 
In both figures the dashed line is merely a guide to the eye.}
\end{figure}

In a second setup, we used a different strategy to demonstrate the appearance 
of moir\'e patterns in quantum images. The spatial profile of the quantum 
correlations between the twin photons carries the combined effect of the 
transmission through two obstacles remotely placed in each photon path 
\cite{padua}. It is useful to interpret this effect in an advanced wave 
picture, where one of the twin photon detectors represents an ideal 
point like source, whose emission is \lq\lq reflected\rq\rq  
at the crystal surface and detected on the other detector \cite{klyshko}. 
Based on this picture we developed the 
experimental setup shown in Fig. 4. The idler detector (D2) is regarded as 
a point like source. A collimating lens (L1) is introduced at a distance 
equal to the focal length $f$ from the \lq\lq source\rq\rq. 
The first grating (G1) is placed at a distance $2f$ from L1. The 
signal beam optics follows the advanced wave reasoning, so that a second 
lens (L2), identical to L1, projects the image of G1 over the second 
grating (G2). All distances are chosen in order to preserve the original 
size of G1 in the projection. Finally, the image of the superposed gratings 
is projected by a third lens (L3, identical to L1 and L2) at the signal 
detection plane, so that the product G1 $\cdot$ G2 is produced in this 
case as well. As in the first setup, we scanned the gratings rather 
than the detectors, since the coincidence region was quite narrow in this 
case as well.
\begin{figure}[ht]
\includegraphics[clip=,width=9cm]{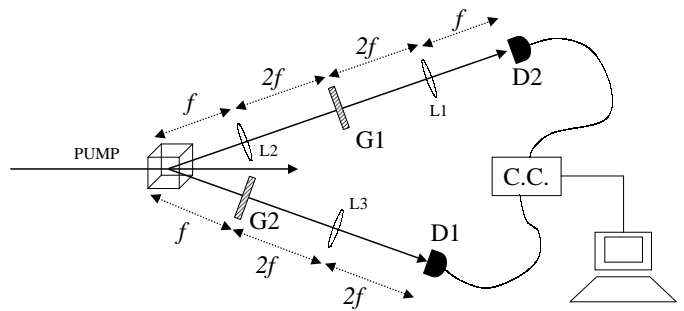}
\caption{\label{setup2}Experimental setup for signal-idler moir\'e}
\end{figure}

The experimental results obtained with the second setup are shown in Fig. 5.
In Fig. 5a, we used the same gratings as in Fig. 1a. The slow modulation 
at a 4.8 mm period is again very clear. The solid line is the product of 
two cosine squared functions and is in very good agreement with the 
experimental data. In Fig. 5b we used the same gratings as in Fig. 1b. 
The slow modulation at 7.8 mm period is also very clear. The solid line is 
a single cosine squared function with this slow period.
\begin{figure}[ht]
\includegraphics[clip=,width=8cm]{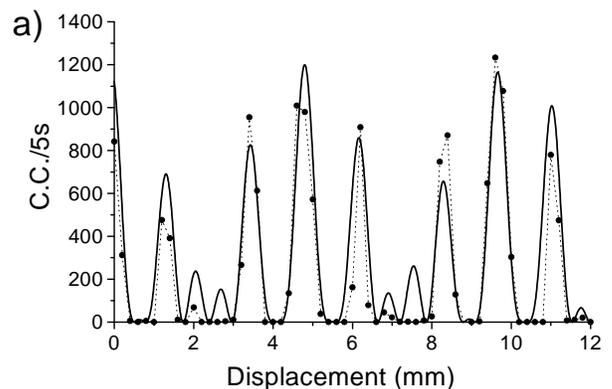}
\includegraphics[clip=,width=8cm]{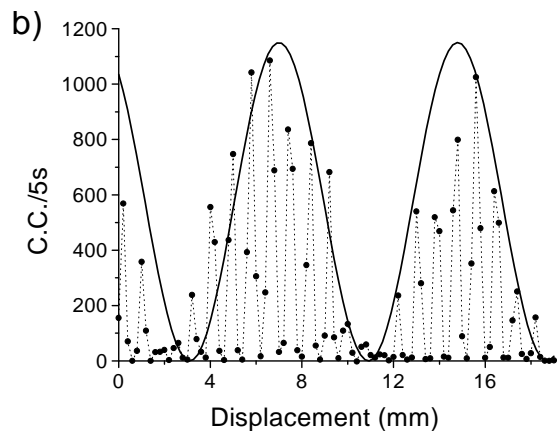}
\caption{\label{result2}Experimental results for signal-idler moir\'e.
a-) 1.2-1.6 mm moir\'e. The solid line is a 
$\cos^2(\pi\,x/1.2mm)\times\cos^2(\pi\,x/1.6mm)$ fit.
b-) 0.8-0.9 mm moir\'e. The solid line is a $\cos^2(\pi\,x/7.8mm)$ fit 
for the slow moir\'e modulation. 
In both figures the dashed line is merely a guide to the eye.}
\end{figure}

In conclusion, we performed the first demonstration of moir\'e patterns 
in quantum images. Two strategies were employed, the first one being 
the transfer of images and angular spectrum from the pump beam to the 
spatial quantum correlations between the twin photons. The second 
strategy made use of an advanced wave picture to produce the moir\'e
patterns with two similar gratings, remotely placed in the signal and 
idler beams. Our results may motivate interesting applications to 
noncontact measurements of small mechanical deformations. 
The moir\'e patterns appear in a variety of different fields, where 
these results could be of great interest.

\begin{acknowledgments}
The authors thank P.A.M. dos Santos for fruitful discussions.
This work is supported by the Conselho Nacional de 
Desenvolvimento Cient\'{\i}fico e Tecnol\'ogico (CNPq) through the 
{\it Instituto do Mil\^enio de Informa\c c\~ao Qu\^antica} 
and {\it Programa de N\'ucleos de Excel\^encia} (PRONEX). 
The PRONEX project is also supported by the Funda\c c\~{a}o de Amparo 
\`{a} Pesquisa do Estado do Rio de Janeiro (FAPERJ).
The authors aknowledge partial funding from Coordena\c c\~{a}o de 
Aperfei\c coamento de Pessoal de N\' \i vel Superior 
(CAPES/PROCAD and CAPES/COFECUB projects).
\end{acknowledgments}

\end{document}